\documentstyle[aps]{revtex}

\begin{document}
\title{Exploiting Particle Statistics in Quantum Computation}
\author{Giuseppe Castagnoli}
\address{Information Technology Dept., Elsag Bailey, 16154 Genova, Italy}
\author{Dalida Monti}
\address{Universit\`{a} di Genova and Elsag Bailey, 16154 Genova, Italy}
\date{\today }
\maketitle

\begin{abstract}
We describe a plausible-speculative form of quantum computation which
exploits particle (fermionic, bosonic) statistics, under a generalized,
counterfactual interpretation thereof. In the idealized situation of an
isolated system, it seems that this form of computation yields to
NP-complete=P.
\end{abstract}

\section{Introduction}

Particle (fermionic, bosonic) statistics has never been applied until now in
the usual (most fruitful) approach to quantum computation. We mean
sequential computation, which can be characterized by the fact that a
reversible Boolean network appears in the{\em \ time-diagram} of the
computation process$^{\left[ 1-6\right] }$.

It has been applied instead in the approach expounded in ref. [7-12], which
differs from the usual one:

\begin{itemize}
\item  a different form of computation is used; its classical counterpart is
simulated annealing or ground state computation; time is orthogonal to the
layout of a reversible Boolean network whose input and output qubits coexist
as simultaneous eigenvalues of a set of compatible observables. Therefore
part of the network input and part of the output can be both constrained.
Checking whether this network is satisfiable is a version of the NP-complete
SAT problem.

\item  This framework seems suitable to applying particle statistics: the
network gates and wires establish Boolean relations between the simultaneous
eigenvalues of the respective input/output qubits; a particle statistics
symmetry is also a constraint applying to simultaneous eigenvalues.

\item  However, in the conventional interpretation, a particle statistics
symmetry is an initial condition which is conserved as a constant of motion
by a unitary evolution. Therefore it does nothing along the course of this
type of evolution. In the generalized interpretation propounded in ref.
[8-12], a particle statistics symmetry is the result of a {\em continuous} 
{\em projection} on the Hilbert subspace satisfying the symmetry. We can
thus apply a Hamiltonian $G_{r}$ to a part $r$ of the system, provided that
particle indistinguishability is preserved. This {\em and} the continuous
projection, seen as a form of ``interaction'' between $r$ and the remaining
part of the system $s$, define a system level Hamiltonian $G_{rs}$. $G_{rs}$
induces a unitary evolution of the state of the whole system which does not
violate the symmetry. In a way, this evolution is {\em driven} by $G_{r}$
and shaped by particle statistics seen as continuous projection.
\end{itemize}

In [12] we have shown how to map the logical constraint established by a
network element (N.E., namely gate or wire), on a constraint induced by
particle statistics. This constraint is seen as continuous projection on the
Hilbert subspace satisfying the N.E. logical constraint. The network is
prepared in an initial state satisfying the (partial) input constraints and
all N.E. constraints, whereas the output constraint (it is sufficient to
constrain only one output qubit) is temporarily removed. This amounts to
solve a problem polynomial in network size. Then we operate on the (to be
constrained) output qubit in order to bring it in match with its constraint.
This operation, under the continuous projection, induces a unitary
transformation leading the network state to satisfying all the constraints
(with very high probability), if the network is satisfiable. This requires a
time independent of network size and would yield to NP-complete=P, see [12].

However, in [12] we could not exclude the possibility that the operation
performed on the output qubit pushed the network into an error state, with
some probability; the rate of growth of this probability with network size
was not known. An exponential growth would have completely vanified the
former result.

In the current work, we show that this probability can in principle be zero,
which brings to NP-complete=P in an idealized framework.

To show this, we will need to recover part of [12], namely the
counterfactual interpretation of particle statistics and its application to
an elementary network element (a NOT\ gate).

\section{A counterfactual, generalized interpretation of particle statistics}

The first part of this Section summarizes a corresponding part of [12]. The
second part introduces a further development.

Let us consider the unitary evolution of a triplet state induced, for
assumption, by particle statistics:

\begin{eqnarray}
\left| \Psi \left( t\right) \right\rangle &=&\cos ^{2}\left( \vartheta
+\omega t\right) \left| 0\right\rangle _{1}\left| 0\right\rangle _{2}+\sin
\left( \vartheta +\omega t\right) \cos \left( \vartheta +\omega t\right)
\left( \left| 0\right\rangle _{1}\left| 1\right\rangle _{2}+\left|
1\right\rangle _{1}\left| 0\right\rangle _{2}\right) \\
&&+\sin ^{2}\left( \vartheta +\omega t\right) \left| 1\right\rangle
_{1}\left| 1\right\rangle _{2}  \nonumber
\end{eqnarray}
1 and 2 label two identical two-state particles with Hilbert spaces
respectively ${\cal H}_{1}=span\{\left| 0\right\rangle _{1},\left|
1\right\rangle _{1}\}$ and ${\cal H}_{2}=span\{\left| 0\right\rangle
_{2},\left| 1\right\rangle _{2}\}$ (which makes this an idealized situation).

Evolution (1) is obtained by applying an identical rotation $\omega t$ of
the states of the two particles to the initial state $\left| \Psi \left(
0\right) \right\rangle $, namely: $Q_{1}\left( \omega t\right) Q_{2}\left(
\omega t\right) \left| \Psi \left( 0\right) \right\rangle =\left| \Psi
\left( t\right) \right\rangle ,$ where $Q_{j}\left( \omega t\right) =\left( 
\begin{tabular}{cc}
$\cos \omega t$ & $\sin \omega t$ \\ 
$-\sin \omega t$ & $\cos \omega t$%
\end{tabular}
\right) _{j},$ with $j=1,2.$

In the usual interpretation, the propagator $Q_{1}\left( \omega t\right)
Q_{2}\left( \omega t\right) $, symmetrical for the particle permutation $%
P_{12}$, commutes with $S_{12}=\frac{1}{2}\left( 1+P_{12}\right) $;
therefore the triplet state symmetry is an initial condition conserved as a
constant of motion.

The counterfactual interpretation given in [12] can be summarized as
follows. We consider the {\em possibility} that $\left| \Psi \left( \tau
\right) \right\rangle $ at any time $\tau =t-dt$ goes out of symmetry. This
is interpreted by saying that $\left| \Psi \left( t-dt\right) \right\rangle $
is not constrained by the symmetry: it is therefore a free vector of the
Hilbert space ${\cal H}_{12}$ $={\cal H}_{1}\otimes {\cal H}_{2}$, namely:

\begin{eqnarray}
\left| \Psi \left( t-dt\right) \right\rangle &=&\alpha _{00}^{\left(
t-dt\right) }\left| 0\right\rangle _{1}\left| 0\right\rangle _{2}+\frac{1}{2}%
\left( \alpha _{01}^{\left( t-dt\right) }+\alpha _{10}^{\left( t-dt\right)
}\right) \left( \left| 0\right\rangle _{1}\left| 1\right\rangle _{2}+\left|
1\right\rangle _{1}\left| 0\right\rangle _{2}\right) + \\
&&\frac{1}{2}\left( \alpha _{01}^{\left( t-dt\right) }-\alpha _{10}^{\left(
t-dt\right) }\right) \left( \left| 0\right\rangle _{1}\left| 1\right\rangle
_{2}-\left| 1\right\rangle _{1}\left| 0\right\rangle _{2}\right) +\alpha
_{11}^{\left( t-dt\right) }\left| 1\right\rangle _{1}\left| 1\right\rangle
_{2};  \nonumber
\end{eqnarray}

\noindent the $\alpha _{ij}^{\left( t-dt\right) }$\ are free complex
variables independent of each other up to $\sum_{i,j\in \left\{ 0,1\right\}
}\left| \alpha _{ij}^{\left( t-dt\right) }\right| ^{2}=1,$ for any time $t$.
In other words, if $t_{1}\neq t_{2}$, $\left| \Psi \left( t_{1}\right)
\right\rangle $ and $\left| \Psi \left( t_{2}\right) \right\rangle $ are two 
{\em independent} free normalized vectors of ${\cal H}_{12}$.

Of course we must assume the possibility that $\alpha _{01}^{\left(
t-dt\right) }-\alpha _{10}^{\left( t-dt\right) }\neq 0$. If this is the
case, namely if $\left| \Psi \left( t-dt\right) \right\rangle $ is out of
symmetry, this state should be ``immediately'' projected on the symmetric
subspace

\[
{\cal H}_{12}^{\left( s\right) }=span\left\{ \left| 0\right\rangle
_{1}\left| 0\right\rangle _{2},\frac{1}{\sqrt{2}}\left( \left|
0\right\rangle _{1}\left| 1\right\rangle _{2}+\left| 1\right\rangle
_{1}\left| 0\right\rangle _{2}\right) ,\left| 1\right\rangle _{1}\left|
1\right\rangle _{2}\right\} . 
\]

The result of this projection can be obtained by submitting $\left| \Psi
\left( t\right) \right\rangle $, another free normalized vector of ${\cal H}%
_{12}$ independent of $\left| \Psi \left( t-dt\right) \right\rangle $, to
the following equations:

for all $t$:

\begin{enumerate}
\item[i)]  $S_{12}\left| \Psi \left( t\right) \right\rangle =\left| \Psi
\left( t\right) \right\rangle ,$

\item[ii)]  $Max$ $\left| \left\langle \Psi \left( t\right) \right. \left| \
\Psi \left( t-dt\right) \right\rangle \right| ,$ which naturally means that
the distance between the vector before projection $\left| \Psi \left(
t-dt\right) \right\rangle $ and the vector after projection $\left| \Psi
\left( t\right) \right\rangle $ is minimum,

where $\left| \Psi \left( t\right) \right\rangle $ is a free normalized
vector of ${\cal H}_{12}$.
\end{enumerate}

\noindent These equations also take into account the fact that projection
goes on repeatedly in a continuous fashion.

In this way, the triplet state symmetry is seen as the result of a watchdog
effect that continuously projects the system state on the symmetrical
subspace ${\cal H}_{12}^{\left( s\right) }$. Of course the result is that $%
\left| \Psi \left( t\right) \right\rangle $ can never go out of symmetry.
However, this counterfactual reasoning will have consequences (the
possibility that counterfactuals yields to effective consequences in the
quantum framework has been highlighted by R. Penrose$^{[13]}$).

A first consequence is that we do not need to assume that symmetry is an
initial condition: no state satisfying (i) for all $t$ can {\em ever} be out
of symmetry.

Also the notion that symmetry is a constant of motion can be given up. To
show this, it is useful to see the continuous projection (i) and (ii) as a
special, continuous form of state vector reduction. As a matter of fact, the
cancelation of the amplitude $\alpha _{01}^{\left( t-dt\right) }-\alpha
_{10}^{\left( t-dt\right) }$ and the renormalization of the other amplitudes
in state (2) is a partial state vector reduction on the subspace ${\cal H}%
_{12}^{\left( s\right) }$ (it is partial since this subspace has dimension
higher than one). Of course there is no dynamics in state vector reduction,
only interference -- in the above form of cancelation of one amplitude and
renormalization of the others.

We should note the peculiarity that such a reduction always occurs on the
symmetrical subspace, never on the orthogonal subspace. This is paradoxical
in counterfactual reasoning, it would mean that the result of reduction is
not random but is affected by a condition placed in the immediate future,
namely that the result does not violate the symmetry. As a matter of fact,
the current counterfactual interpretation can be justified in a two-way
(advanced and retarded) propagation model$^{\left[ 9\right] }$.

From another standpoint, there is no paradox at all since, actually, $\left|
\Psi \left( t\right) \right\rangle $ never goes out of the subspace ${\cal H}%
_{12}^{\left( s\right) }$. \ However, the consequences of this
counterfactual reasoning {\em will} {\em diverge} from the conventional way
of applying quantum mechanics.

We shall now reconstruct evolution (1) by resorting to conditions (i) and
(ii). A first observation is that these conditions affect the overall state
of the two particles, therefore the states of the individual particles may
no more be defined. We must use the particle density matrices

\[
\rho _{i}\left( t\right) =Tr_{3-i}\left[ \left| \Psi \left( t\right)
\right\rangle \left\langle \Psi \left( t\right) \right| \right] , 
\]

\noindent where $Tr_{3-i}$ means partial trace over $3-i$, and $i=1,2$ is
the particle label (if we use the method of random phases$^{\left[ 14\right]
}$, $\left| \Psi \left( t\right) \right\rangle $ does not need to be a pure
state -- anyhow it will turn out to be that).

A second observation is that the coherence elements of each density matrix
-- as entanglement -- can also be affected by the watchdog effect (in fact
they will be determined by it). We only know that the diagonal of each
density matrix must show an $\omega t$ rotation [to be consistent with eq.
(1)]. In conclusion we must add to conditions (i) and (ii) the further
double condition (iii):

\[
diag\rho _{1}\left( t\right) =diag\left\{ Tr_{2}\left[ \left| \Psi \left(
t\right) \right\rangle \left\langle \Psi \left( t\right) \right| \right]
\right\} =\cos ^{2}\left( \vartheta +\omega t\right) \left| 0\right\rangle
_{1}\left\langle 0\right| _{1}+\sin ^{2}\left( \vartheta +\omega t\right)
\left| 1\right\rangle _{1}\left\langle 1\right| _{1}. 
\]

\[
diag\rho _{2}\left( t\right) =diag\left\{ Tr_{1}\left[ \left| \Psi \left(
t\right) \right\rangle \left\langle \Psi \left( t\right) \right| \right]
\right\} =\cos ^{2}\left( \vartheta +\omega t\right) \left| 0\right\rangle
_{2}\left\langle 0\right| _{2}+\sin ^{2}\left( \vartheta +\omega t\right)
\left| 1\right\rangle _{2}\left\langle 1\right| _{2}. 
\]

It is readily seen that the {\em simultaneous} application of conditions
(i), (ii) and (iii) yields evolution (1). It can be said that evolution (1)
is {\em driven} by conditions (iii) and {\em shaped} by conditions (i) and
(ii)\footnote{%
An example of a unitary evolution shaped by a continuous form of state
vector reduction, is the evolution of the polarization of a photon going
through an infinite series of polarizing filters, each rotated by an
infinitesimal constant amount with respect to the former one. In a way, we
go back to the root of quantum computation (computation reversibility$^{%
\left[ 15,16\right] }$) and take an alternative branch, by exploring a
strictly quantum form of reversible computation.}.

\noindent We should now note a fact which is {\em essential} to the current
work. As readily seen, {\em by removing either one of the two conditions }%
(iii), evolution (1) is still obtained -- the two conditions are redundant
with respect to one another.

Therefore it is perfectly legitimate to say that the {\em rotation of the
state of only} {\em one particle}, either one in an indistinguishable way, 
{\em drags an identical rotation of the state of the other particle}. In
this idealized picture, particle statistics can be seen as an interaction
free constraint, namely as a non-dynamic constraint operating by way of
destructive interference and renormalization.

We will show how to apply (speculatively) the driving condition (iii). This
is a further development with respect to [12]. Only one particle $j$, either
one in an indistinguishable way, should be submitted to the Hamiltonian 
\begin{equation}
G_{j}=\omega \sigma _{yj}{\bf 1}_{3-j},
\end{equation}
where $\sigma _{yj}$ is the Pauli matrix in ${\cal H}_{j}$

\begin{equation}
\sigma _{yj}=\left( 
\begin{tabular}{cc}
$0$ & $i$ \\ 
$-i$ & $0$%
\end{tabular}
\right) _{j},
\end{equation}

\noindent and ${\bf 1}_{3-j}$ is the identity in ${\cal H}_{3-j}$. The
overall Hamiltonian $G_{12}$, operating in ${\cal H}_{12}$, is obtained
through symmetrization of $G_{j}$: $G_{12}=\frac{1}{2}\omega \left( \sigma
_{y1}{\bf 1}_{2}+{\bf 1}_{1}\sigma _{y2}\right) $. This yields

\begin{equation}
G_{12}=\frac{1}{2}\omega \left( 
\begin{tabular}{cccc}
$0$ & $i$ & $i$ & $0$ \\ 
$-i$ & $0$ & $0$ & $i$ \\ 
$-i$ & $0$ & $0$ & $i$ \\ 
$0$ & $-i$ & $-i$ & $0$%
\end{tabular}
\right) .
\end{equation}

\noindent This is in fact the generator of $Q_{1}\left( \omega t\right)
Q_{2}\left( \omega t\right) $, as readily checked.

The above symmetrization is interpreted as follows. If we applied the
Hamiltonian $G_{j}$ to one particle assumed to be independent of the other,
we would have obtained a rotation of the state of that particle by $\omega t$%
. By considering the continuous projection on ${\cal H}_{12}^{\left[ s\right]
}$ as a form of interdependence between the two particles, the former
Hamiltonian becomes $G_{12}$ at overall system level (in a way, continuous
projection is considered as a sort of interaction Hamiltonian between the
two particles). $G_{12}$ generates the operator $Q_{1}\left( \omega t\right)
Q_{2}\left( \omega t\right) $ which rotates the states of both particles of
the same amount. Therefore we can say that the rotation of the state of only
one particle (either one in an indistinguishable way), drags an identical
rotation of the state of the other.

This, in the current context, is a tautological interpretation of particle
statistics. However, applied to a different context (this will be two
indistinguishable particles hosted by two distinguishable lattice sites),
such an interpretation will yield two results diverging from the
conventional way of applying quantum mechanics.

\section{An elementary gate as a projector}

We consider a couple of (coexisting) qubits $r$ and $s$ which make up the
input and the output of a NOT\ gate. Let

\[
{\cal H}_{rs}=span\left\{ \left| 0\right\rangle _{r}\left| 0\right\rangle
_{s},\left| 0\right\rangle _{r}\left| 1\right\rangle _{s},\left|
1\right\rangle _{r}\left| 0\right\rangle _{s},\left| 1\right\rangle
_{r}\left| 1\right\rangle _{s}\right\} , 
\]

\noindent be the Hilbert space of the two qubits, 
\[
{\cal H}_{rs}^{\left( c\right) }=span\left\{ \left| 0\right\rangle
_{r}\left| 1\right\rangle _{s},\left| 1\right\rangle _{r}\left|
0\right\rangle _{s}\right\} 
\]

\noindent be the constrained subspace, spanned by those ${\cal H}_{rs}$
basis vectors which satisfy the NOT\ gate, and $A_{rs}$ be the projector
from ${\cal H}_{rs}$ on ${\cal H}_{rs}^{\left( c\right) }$.

We shall apply the mathematical model of Section II to represent an
evolution of the state of the NOT\ gate $\left| \Psi \left( t\right)
\right\rangle $\ (this Section is purely mathematical, a plausible physical
model will be given in Section IV). $\left| \Psi \left( t\right)
\right\rangle $ should be a free normalized vector of ${\cal H}_{rs}$, 
\[
\left| \Psi \left( t\right) \right\rangle =\sum_{i,j\in \left\{ 0,1\right\}
}\alpha _{ij}^{\left( t\right) }\left| i\right\rangle _{r}\left|
j\right\rangle _{s},\text{ with}\sum_{i,j\in \left\{ 0,1\right\} }\left|
\alpha _{ij}^{\left( t\right) }\right| ^{2}=1, 
\]
subject to continuous projection on ${\cal H}_{rs}^{\left( c\right) }$:

\bigskip for all $t$:

\begin{enumerate}
\item[i)]  \smallskip $A_{rs}\left| \Psi \left( t\right) \right\rangle
=\left| \Psi \left( t\right) \right\rangle ,$

\item[ii)]  Max $\left| \left\langle \Psi \left( t\right) \right. \left|
\Psi \left( t-dt\right) \right\rangle \right| .$
\end{enumerate}

Let $\left| \Psi \left( 0\right) \right\rangle =\cos \vartheta \left|
0\right\rangle _{r}\left| 1\right\rangle _{s}+\sin \vartheta \left|
1\right\rangle _{r}\left| 0\right\rangle _{s}$ be the gate initial state. We
assume of acting on qubit $r$ with the Hamiltonian

\[
G_{r}=\omega \sigma _{yr},\text{ where }\sigma _{yr}=\omega \left( 
\begin{tabular}{cc}
$0$ & $i$ \\ 
$-i$ & $0$%
\end{tabular}
\right) _{r}. 
\]

We have two ways of deriving the evolution induced by $G_{r}$.

a) If acting on the independent qubit $r$, $G_{r}$ would rotate its state by 
$\omega t$. As we have seen in Section II, the diagonal of qubit $r$ density
matrix should not be affected by the continuous projection on ${\cal H}%
_{rs}^{\left( c\right) }$. Thus:

\[
\text{iii) }diag\rho _{r}\left( t\right) =diag\left\{ Tr_{s}\left[ \left|
\Psi \left( t\right) \right\rangle \left\langle \Psi \left( t\right) \right| %
\right] \right\} =\cos ^{2}\left( \vartheta +\omega t\right) \left|
0\right\rangle _{r}\left\langle 0\right| _{r}+\sin ^{2}\left( \vartheta
+\omega t\right) \left| 1\right\rangle _{r}\left\langle 1\right| _{r}, 
\]

Conditions (i), (ii) and (iii) define the unitary evolution

\begin{equation}
\left| \Psi \left( t\right) \right\rangle =\cos \left( \vartheta +\omega
t\right) \left| 0\right\rangle _{r}\left| 1\right\rangle _{s}+\sin \left(
\vartheta +\omega t\right) \left| 1\right\rangle _{r}\left| 0\right\rangle
_{s},
\end{equation}

\noindent as readily checked. Condition (iii) {\em drives} and conditions
(i) and (ii) {\em shape} this evolution. We can see that the rotation of
qubit $r$ [the driving condition $G_{r}$, or (iii)] induces an identical
rotation of qubit $s$:

\[
\rho _{s}\left( t\right) =T_{r_{r}}\left[ \left| \Psi \left( t\right)
\right\rangle \left\langle \Psi \left( t\right) \right| \right] =\sin
^{2}\left( \vartheta +\omega t\right) \left| 0\right\rangle _{s}\left\langle
0\right| _{s}+\cos ^{2}\left( \vartheta +\omega t\right) \left|
1\right\rangle _{s}\left\langle 1\right| _{s}, 
\]

\noindent of course 0 and 1 are interchanged.

b) A second way of deriving evolution (6) consists in computing the
Hamiltonian $G_{rs}$ acting on the overall state of the two qubits. $G_{rs}$
is originated by $G_{r}$ {\em and} the continuous projection (seen as a form
of interaction or better interdependence between the two qubits). Given that
this latter introduces the following bijective correspondence between ${\cal %
H}_{r}$ and ${\cal H}_{rs}$: 
\[
\left| 0\right\rangle _{r}\leftrightarrow \left| 0\right\rangle _{r}\left|
1\right\rangle _{s},\quad \left| 1\right\rangle _{r}\leftrightarrow \left|
1\right\rangle _{r}\left| 0\right\rangle _{s}, 
\]

\noindent $G_{r}$ is:

\[
G_{rs}=\left( 
\begin{tabular}{cccc}
$0$ & $0$ & $0$ & $0$ \\ 
$0$ & $0$ & $i$ & $0$ \\ 
$0$ & $-i$ & $0$ & $0$ \\ 
$0$ & $0$ & $0$ & $0$%
\end{tabular}
\right) ; 
\]

\noindent naturally $G_{r}$ populates the central part of $G_{rs}$.

As readily checked, $G_{rs}$ generates the operator:

$Q_{rs}\left( \omega t\right) \equiv \left( 
\begin{array}{cccc}
1 & 0 & 0 & 0 \\ 
0 & \cos \omega t & \sin \omega t & 0 \\ 
0 & -\sin \omega t & \cos \omega t & 0 \\ 
0 & 0 & 0 & 1
\end{array}
\right) $,

with $\left| 0\right\rangle _{r}\left| 1\right\rangle _{s}\equiv \left( 
\begin{array}{l}
1 \\ 
0
\end{array}
\right) _{r}\otimes \left( 
\begin{array}{l}
0 \\ 
1
\end{array}
\right) _{s},\left| 1\right\rangle _{r}\left| 0\right\rangle _{s}\equiv
\left( 
\begin{array}{l}
0 \\ 
1
\end{array}
\right) _{r}\otimes \left( 
\begin{array}{l}
1 \\ 
0
\end{array}
\right) _{s}.$

One can see that $Q_{rs}\left( \omega t\right) \left| \Psi \left( 0\right)
\right\rangle =\left| \Psi \left( t\right) \right\rangle $ given by equation
(6). We have thus obtained the same result as before -- when the driving
condition was the evolution of $diag\rho _{r}\left( t\right) $.

In view of what will follow, it is important to note that, if all ${\cal H}%
_{rs}^{\left( c\right) }$ basis vectors occur with amplitudes different from
zero in the initial state ($t=0$), namely if $\vartheta \neq 0,\frac{\pi }{2}
$, condition (i) is redundant with respect to condition (ii). In this case,
condition (ii) alone implies $\alpha _{00}^{\left( t\right) }=\alpha
_{11}^{\left( t\right) }=0$ which already satisfies condition (i).

On the contrary, condition (i) is not redundant if $\vartheta =0$ or $\frac{%
\pi }{2}$. For example, if $\vartheta =0$, i.e. $\left| \Psi \left( 0\right)
\right\rangle =\left| 0\right\rangle _{r}\left| 1\right\rangle _{s}$,
condition (ii) implies $\alpha _{00}^{\left( t\right) }=0$ and $\left|
\alpha _{01}^{\left( t\right) }\right| ^{2}=\cos ^{2}\left( \vartheta
+\omega t\right) $, as needed, while $\alpha _{10}^{\left( t\right) }$ and $%
\alpha _{11}^{\left( t\right) }$ are only subject to the constraint $\left|
\alpha _{10}^{\left( t\right) }\right| ^{2}+\left| \alpha _{11}^{\left(
t\right) }\right| ^{2}=\sin ^{2}\left( \vartheta +\omega t\right) $. Thus,
disregarding condition (i) would allow for the existence of the forbidden
state $\alpha _{11}^{\left( t\right) }\left| 1\right\rangle _{r}\left|
1\right\rangle _{s}$.

In conclusion, we have ascertained a peculiar fact. Our ``operation on a
part'' [this is just the mathematical condition (iii), or $G_{rs}$ in
equivalent terms, for the time being], {\em blind} to its effect on the
whole, performed together with continuous $A_{rs}$ projection, generates a 
{\em unitary} {\em transformation} which is, so to speak, {\em wise} to the
whole state, to how it should be transformed without ever violating $A_{rs}$
(i.e. the NOT\ gate). Of course $A_{rs}$ ends up commuting with the
resulting overall unitary propagator, but because this is itself {\em shaped}
by $A_{rs}$.

\section{Exploiting particle statistics}

$A_{rs}$ projection can be shown to be an epiphenomenon of particle
(fermionic or bosonic) statistics ``turned on'' in a special physical
situation. In the following, we will adopt fermionic statistics.

In order to implement the NOT\ gate, we consider two identical fermionic
particles 1 and 2. Just for the sake of visualization, we can think that
each particle has spin 1/2 and can occupy either one of two distinguishable
lattice sites $r$ and $s$. \ The generic basis vector of this system has the
form $\left| \chi _{1}\right\rangle _{1}\left| \chi _{2}\right\rangle
_{2}\left| \lambda _{1}\right\rangle _{1}\left| \lambda _{2}\right\rangle
_{2}$, where $\chi _{i}=0,1$ means that the spin of particle $i$ is down, up
and $\lambda _{i}=r,s$ means that the site occupied by particle $i$ is $r,$ $%
s$. For example, $\left| 0\right\rangle _{1}\left| 1\right\rangle _{2}\left|
r\right\rangle _{1}\left| s\right\rangle _{2}$ reads: particle $1$ spin $=$
0, particle $2$ spin $=$ 1, particle $1$ site $=r,$ particle $2$ site $=s$.

16 combinations like this make up the basis of the Hilbert space ${\cal H}%
_{12}$. However, there are only six {\em antisymmetrical} combinations (not
violating fermion statistics) which make up the basis of the {\em %
antisymmetrical} subspace ${\cal H}_{12}^{\left( a\right) }$.

These basis vectors are represented in second quantization and, when there
is exactly one particle per site, in qubit notation ($\chi $ and $\lambda $
stand respectively for the qubit eigenvalue and label), $\left|
0\right\rangle $ is the vacuum vector:

$\left| a\right\rangle =a_{0r}^{\dagger }\ a_{1r}^{\dagger }\left|
0\right\rangle ,$

$\left| b\right\rangle =a_{0s}^{\dagger }\ a_{1s}^{\dagger }\left|
0\right\rangle ;$

$\left| c\right\rangle =a_{0r}^{\dagger }\ a_{0s}^{\dagger }\left|
0\right\rangle =\left| 0\right\rangle _{r}\left| 0\right\rangle _{s},$

$\left| d\right\rangle =a_{1r}^{\dagger }\ a_{1s}^{\dagger }\left|
0\right\rangle =\left| 1\right\rangle _{r}\left| 1\right\rangle _{s},$

$\left| e\right\rangle =\frac{1}{\sqrt{2}}\left( a_{0r}^{\dagger }\
a_{1s}^{\dagger }+a_{1r}^{\dagger }\ a_{0s}^{\dagger }\right) \left|
0\right\rangle =\frac{1}{\sqrt{2}}\left( \left| 0\right\rangle _{r}\left|
1\right\rangle _{s}+\left| 1\right\rangle _{r}\left| 0\right\rangle
_{s}\right) .$

$\left| f\right\rangle =\frac{1}{\sqrt{2}}\left( a_{0r}^{\dagger }\
a_{1s}^{\dagger }-a_{1r}^{\dagger }\ a_{0s}^{\dagger }\right) \left|
0\right\rangle =\frac{1}{\sqrt{2}}\left( \left| 0\right\rangle _{r}\left|
1\right\rangle _{s}-\left| 1\right\rangle _{r}\left| 0\right\rangle
_{s}\right) .$

\noindent $a_{\chi \lambda }^{\dagger }$ creates a particle of spin $\chi $
in site $\lambda $; creation/annihilation operators are subject to: $\left\{
a_{i}^{\dagger },a_{j}^{\dagger }\right\} =\left\{ a_{i},a_{j}\right\} =0,\
\ \left\{ a_{i}^{\dagger },a_{j}\right\} =\delta _{i,j}.$ Under the
condition that there is exactly one particle per site, they generate a {\em %
qubit algebra}.

Now we introduce the Hamiltonian

\[
H_{rs}=-(E_{a}\ a_{0r}^{\dagger }\ a_{1r}^{\dagger }a_{0r}a_{1r}+E_{b}\
a_{0s}^{\dagger }\ a_{1s}^{\dagger }a_{0s}a_{1s}+E_{c}\ a_{0r}^{\dagger }\
a_{0s}^{\dagger }a_{0r}a_{0s}+E_{d}\ a_{1r}^{\dagger }\ a_{1s}^{\dagger
}a_{1r}a_{1s}), 
\]
with $E_{a}$, $E_{b}$, $E_{c}$, $E_{d}\geq E$ discretely above 0. This
leaves us with two degenerate ground eigenstates:

\[
\left| e\right\rangle =\frac{1}{\sqrt{2}}\left( \left| 0\right\rangle
_{r}\left| 1\right\rangle _{s}+\left| 1\right\rangle _{r}\left|
0\right\rangle _{s}\right) \text{ and }\left| f\right\rangle =\frac{1}{\sqrt{%
2}}\left( \left| 0\right\rangle _{r}\left| 1\right\rangle _{s}-\left|
1\right\rangle _{r}\left| 0\right\rangle _{s}\right) . 
\]
The generic ground state is thus:

\begin{equation}
\left| \Psi \right\rangle =\alpha \left| 0\right\rangle _{r}\left|
1\right\rangle _{s}+\beta \left| 1\right\rangle _{r}\left| 0\right\rangle
_{s}\text{, with }\left| \alpha \right| ^{2}+\left| \beta \right| ^{2}=1.
\end{equation}
Of course $\left| \Psi \right\rangle $ satisfies $A_{rs}\left| \Psi
\right\rangle =\left| \Psi \right\rangle $, and belongs to ${\cal H}%
_{rs}^{\left( c\right) }$ (Section III), a subspace of ${\cal H}%
_{12}^{\left( a\right) }$.

Let $A_{12}\left| \Psi \right\rangle =\frac{1}{2}$ $\left( 1-P_{12}\right) $
be the usual antisymmetrization projector. Due to the above
anticommutation~relations: 
\[
A_{12}\left| 0\right\rangle _{r}\left| 1\right\rangle _{s}=\left|
0\right\rangle _{r}\left| 1\right\rangle _{s}\text{ and }A_{12}\left|
1\right\rangle _{r}\left| 0\right\rangle _{s}=\left| 1\right\rangle
_{r}\left| 0\right\rangle _{s},\text{ moreover} 
\]
\[
A_{12}\left| 0\right\rangle _{r}\left| 0\right\rangle _{s}=\left|
0\right\rangle _{r}\left| 0\right\rangle _{s}\text{ and }A_{12}\left|
1\right\rangle _{r}\left| 1\right\rangle _{s}=\left| 1\right\rangle
_{r}\left| 1\right\rangle _{s}, 
\]
without forgetting that $\left| 0\right\rangle _{r}\left| 0\right\rangle
_{s}=\left| c\right\rangle $ and $\left| 1\right\rangle _{r}\left|
1\right\rangle _{s}=\left| d\right\rangle $ are {\em excited states}.

The NOT\ gate can be implemented by suitably operating on the ground state
(7). We assume this to be initially:

\begin{equation}
\left| \Psi \left( 0\right) \right\rangle =\cos \vartheta \left|
0\right\rangle _{r}\left| 1\right\rangle _{s}+\sin \vartheta \left|
1\right\rangle _{r}\left| 0\right\rangle _{s}.
\end{equation}

The evolution induced by the Hamiltonian $G_{r}$ is obtained by adopting the
projection interpretation of particle statistics. This means that $\left|
\Psi \left( t\right) \right\rangle $ is continuously projected on the
antisymmetric subspace $H_{12}^{\left( a\right) }$ while $diag$ $\rho
_{r}\left( t\right) $ evolves as if $G_{r}$ were applied to the independent
qubit $r$:

\noindent {\em for all }$t$:

\begin{enumerate}
\item[i)]  $A_{12}\left| \Psi \left( t\right) \right\rangle =\left| \Psi
\left( t\right) \right\rangle ,$

\item[ii)]  $Max$ $\left| \left\langle \Psi \left( t\right) \right| \left.
\Psi \left( t-dt\right) \right\rangle \right| ,$

\item[iii)]  $diag$ $\rho _{r}\left( t\right) =diag\left\{ Tr_{s}\left[
\left( \left| \Psi \left( t\right) \right\rangle \left\langle \Psi \left(
t\right) \right| \right) \right] \right\} =$
\end{enumerate}

$\cos ^{2}\left( \vartheta +\omega t\right) \left| 0\right\rangle
_{r}\left\langle 0\right| _{r}+\sin ^{2}\left( \vartheta +\omega t\right)
\left| 1\right\rangle _{r}\left\langle 1\right| _{r},$

\begin{enumerate}
\item[iv)]  $\left\langle \xi _{rs}\left( t\right) \right\rangle
=\left\langle \Psi \left( t\right) \right| H_{rs}\left| \Psi \left( t\right)
\right\rangle =0,$

\noindent where $\left| \Psi \left( t\right) \right\rangle $ is a free
normalized vector of ${\cal H}_{12}$.
\end{enumerate}

The solution of the above equations is the desired evolution (6), repeated
here for convenience:

\[
\left| \Psi \left( t\right) \right\rangle =\cos \left( \vartheta +\omega
t\right) \left| 0\right\rangle _{r}\left| 1\right\rangle _{s}+\sin \left(
\vartheta +\omega t\right) \left| 1\right\rangle _{r}\left| 0\right\rangle
_{s}. 
\]

\noindent \noindent Conditions (i), (ii) and (iii) mean that the link state
undergoes a transformation [driven by (iii) or $G_{r}$] under continuous
state vector reduction on the {\em antisymmetric} subspace ${\cal H}%
_{12}^{\left( a\right) }$. If $\vartheta \neq 0,\frac{\pi }{2}$, namely if
the preparation (4) comprises {\em all} the basis vectors of ${\cal H}%
_{rs}^{\left( c\right) }$, condition (ii) alone keeps the link evolution
inside ${\cal H}_{rs}^{\left( c\right) }$ (Section III). The link state
remains ground and consequently the link expected energy $\left\langle \xi
_{rs}\left( t\right) \right\rangle $ is always zero. By excluding $\vartheta
=0,\frac{\pi }{2}$, {\em condition (iv) is a} {\em consequence} {\em of the
former conditions}.

Mathematically, conditions (i) and (iv) give the constraint $A_{rs}\left|
\Psi \left( t\right) \right\rangle =\left| \Psi \left( t\right)
\right\rangle $. In conclusion the above conditions (i) through (iv) \
(which imply interpreting fermionic antisymmetry $A_{12}$ as {\em continuous}
{\em projection} on the antisymmetric subspace) are equivalent to condition
(i) through (iii) of Section III. This gives in fact the evolution (6).

Since $\left[ G_{rs\text{ }},H_{rs}\right] =0$, as readily checked, the
application of $G_{r\text{ }}$to qubit $r$ does not disturb in principle the
ground state of $H_{rs}$, in constrast with what hypothesized in ref. [12].
This means NP-complete=P$^{\left[ 12\right] }$ under the current form of
computation. Of course we should keep in mind that we are in the same
idealized and speculative context of the former work.

\section{Conclusions}

By using the generalized interpretation of particle statistics (viewed as
continuous projection on a constrained subspace), we have obtained an
evolution of the NOT gate by acting {\em only} on qubit $r$ (an input or an
output qubit of the gate, indifferently so given that the two coexist) and
without violating the NOT\ gate logical constraint.

This would be impossible under the conventional interpretation of particle
statistics. In this context, since all ${\cal H}_{rs}$ basis vectors are
already antisymmetrical, particle statistics would do nothing. Any operation
performed {\em only} on qubit $r$ in the initial entangled state (8), would
necessarily originate the terms $\left| 0\right\rangle _{r}\left|
0\right\rangle _{s}$ and $\left| 1\right\rangle _{r}\left| 1\right\rangle
_{s}$, thus violating the NOT gate logical constraint.

If the model Hamiltonians used in this work could be substituted by more
concrete Hamiltonians -- say implementable in a laboratory -- in principle
such a divergence between the two interpretations of particle statistics
could be verified.

This research has been partly developed during the Elsag Bailey-ISI Workshop
on Quantum Computation (Turin, 1997). Thanks are due to A. Ekert, D.
Finkelstein, S. Lloyd and V. Vedral for useful suggestions.

\end{document}